\documentclass[a4paper]{jpconf}
\usepackage{graphicx}
\begin{document}
\title{VLBI observations of nearby radio loud \\
Active Galactic Nuclei}

\author{Gabriele Giovannini$^{1,2}$, Elisabetta Liuzzo$^{1,2}$, 
Marcello Giroletti$^1$, Gregory B Taylor$^3$}

\address{$^1$ Istituto di Radioastronomia - INAF, via Gobetti 101, 40129 
Bologna, Italy}
\address{$^2$ Dipartimento di Astronomia, Universita' di Bologna, via Ranzani 
1, 40127 Bologna, Italy}
\address{$^3$ Department of Physics and Astronomy, University of New Mexico, 
Albuquerque NM 87131, USA} 

\ead{ggiovann@ira.inaf.it}

\begin{abstract}
We present an update of the parsec scale properties of the Bologna Complete 
Sample 
consisting of 95 radio sources from the B2 
Catalog of Radio Sources and the Third Cambridge Revised Catalog (3CR), with 
z $<$ 0.1. 
Thanks to recent new data we have now parsec scale images for 76 sources 
of the 
sample. Most of them show a one-sided jet structure but we find a higher 
fraction 
of two-sided sources in comparison with previous flux-limited VLBI 
surveys. A few peculiar sources are presented and discussed in more 
detail.
\end{abstract}

\section{Introduction}

The study of the parsec scale properties of radio galaxies is crucial to
obtain information on the nature of their central engine.

In order to study statistical properties of different classes of sources, it
is necessary to define and observe a sample that is free from selection
effects. To this aim, it is important to define the sample using a low radio
frequency, to avoid observational biases related to orientation effects,
thanks to the dominance at low radio frequency of the unbeamed extended
emission. With this purpose, we started a project to observe a complete
sample of radio galaxies (the Bologna Complete Sample (BCS)) 
selected within the B2 Catalog of Radio Sources and
the Third Cambridge Revised Catalog (3CR) \cite{gio90, gio01}, 
with no selection constrain on the nuclear 
properties.

The sample consists of 95 radio sources. We selected
the sources above a homogeneous flux density limit of 0.25 Jy at 408 MHz for
the B2 sources and those above 10 Jy at 178 MHz for the 3CR sources 
\cite{fer84} and applied the following criteria:

\vskip 0.5truecm

1) declination $>$10$^{\circ}$

2) Galactic latitude $|$ b $|$ $>$ 15 $^{\circ}$

3) redshift z$<$0.1

\vskip 0.5truecm

At present VLBI observations are available for 76 sources. The missing 19
sources will require very sensitive observations (phase reference technique 
and large bandwidth) because of the very low activity of the radio
core.

\section{Source morphology}

Among sources with VLBI data, we have the following kiloparsec-scale 
structure: 

\begin{itemize}

\item
50 FR I radio galaxies; 
\item
13 FR II radio galaxies; 
\item
8 compact sources with a flat spectrum; 
\item
2 BL Lac objects; 
\item
1 compact symmetric object (CSO); 
\item
1 compact steep-spectrum source (CSS). 
\end{itemize}

Moreover, 1 source
is identified with a peculiar Spiral galaxy and was not included
in the present discussion.

The observed sample is representative of a sample of sources oriented 
at random angles with respect 
to the line of sight. In fact, in a randomly oriented sample of radio galaxies,
the probability that a source is at an angle between $\theta_{1}$ and 
$\theta_{2}$ 
with respect to the line of sight is
 P($\theta_{1}$,$\theta_{2}$) = cos$\theta_{1}$ $-$ cos$\theta_{2}$.
We find that the percentage of FR I plus FR II radio sources is $\sim$
$80\%$, corresponding to sources oriented at angles greater than 
$\sim$ 35$^{\circ}$ -- 40$^{\circ}$, in agreement with unified models.

At parsec scale most of the sources (24 FR I and 7 FR II) show as expected a
one-sided jet structure because of relativistic effects, however we have
also a high number of sources with a symmetric jet structure. 
We note, however, that the detection of a counterjet may 
be related to the sensitivity of the map, whereas the dynamic range is not a 
problem in our images, as the sources are generally weak. We classify as two 
sided all sources showing both a jet and counterjet, regardless of the value 
of the jet to the counterjet ratio or the length of the counterjet.

The total number of sources with a two-sided morphology is 17, corresponding 
to $\sim$ 25$\%$. This percentage
 is significantly higher than that found in previous samples in the 
literature:  
there are 11$\%$ symmetric sources in the PR (Pearson-Readhead) sample 
\cite{pea88} and 4.6$\%$ (19/411) in the 
combined PR and 
Caltech-Jodrell (CJ) samples \cite{tay94,pol95}. 

For two-sided sources, the brightness ratio between the two jets is in 
the range 1--5, while most of the brightness lower limit in one-sided
sources is higher than 5 confirming that the 
sensitivity in our maps is generally good enough to detect two sided sources.

The main difference between the percentage of symmetric sources in the 
present sample and in previous samples is naturally explained in the framework
 of unified scheme models since our sources have been 
selected at low frequency and should not be 
affected by an orientation bias.  

The comparison between the VLA and VLBA jet position angles (P.A.)
shows that most of the 
sources do not 
show a large misalignment and that 
the one-sided parsec scale (or the brigther jet in double-sided parsec jets) 
is oriented with respect to the nuclear emission, on the same side of the 
brighter kiloparsec-scale.

We compared the total flux at VLBA scale with the core arcsecond flux density.
We find that 70 $\%$ have a 
correlated flux density larger than 70$\%$ of the arcsecond core flux density.
 This means that in these sources we mapped most of the milliarcsecond (mas) 
scale structure 
and so we can properly connect the parsec to the kiloparsec structures. 

For 19 (30$\%$) sources, a significant fraction of the arcsecond core 
flux density is missing in the VLBA images. This suggest variability or  the 
presence of significant structures between $\sim$30 mas and 2 arcsec that the 
VLBA can miss because the lack of short baselines. 
To properly study these structures, future observations 
with EVLA or the e-MERLIN array will be necessary.

\section{Individual sources}

{\bf 0802+24 -- 3C192}. This source
has an X symmetric double-lobe structure which extends 212 arcsec
at 8.35 GHz (see Fig. 1 \footnote {http://www.jb.man.ac.uk/atlas/}), 
showing bright hotspots at the end of the lobes
\cite{bau88,har98}.
Extended and highly ionized emission lines were detected in this galaxy 
\cite{bau92}.
In our VLBA image (Fig. 2), the source appears two-sided oriented as the
kpc jet (P.A. $\sim$-80$^{\circ}$), with a core flux
density $\sim$2.1 mJy and a jet flux density $\sim$1.45 mJy.

The observed core radio power is very low with respect to the total radio power
(see also Fig. 1)
suggesting the presence of very fast jets at a large angle with respect to
the line of sight and therefore a Doppler factor $<<$ 1. This result is in
agreement with the presence of faint symmetric jets in our VLBA image.

\begin{figure}[h]
\begin{minipage}{14pc}
\includegraphics[width=14pc]{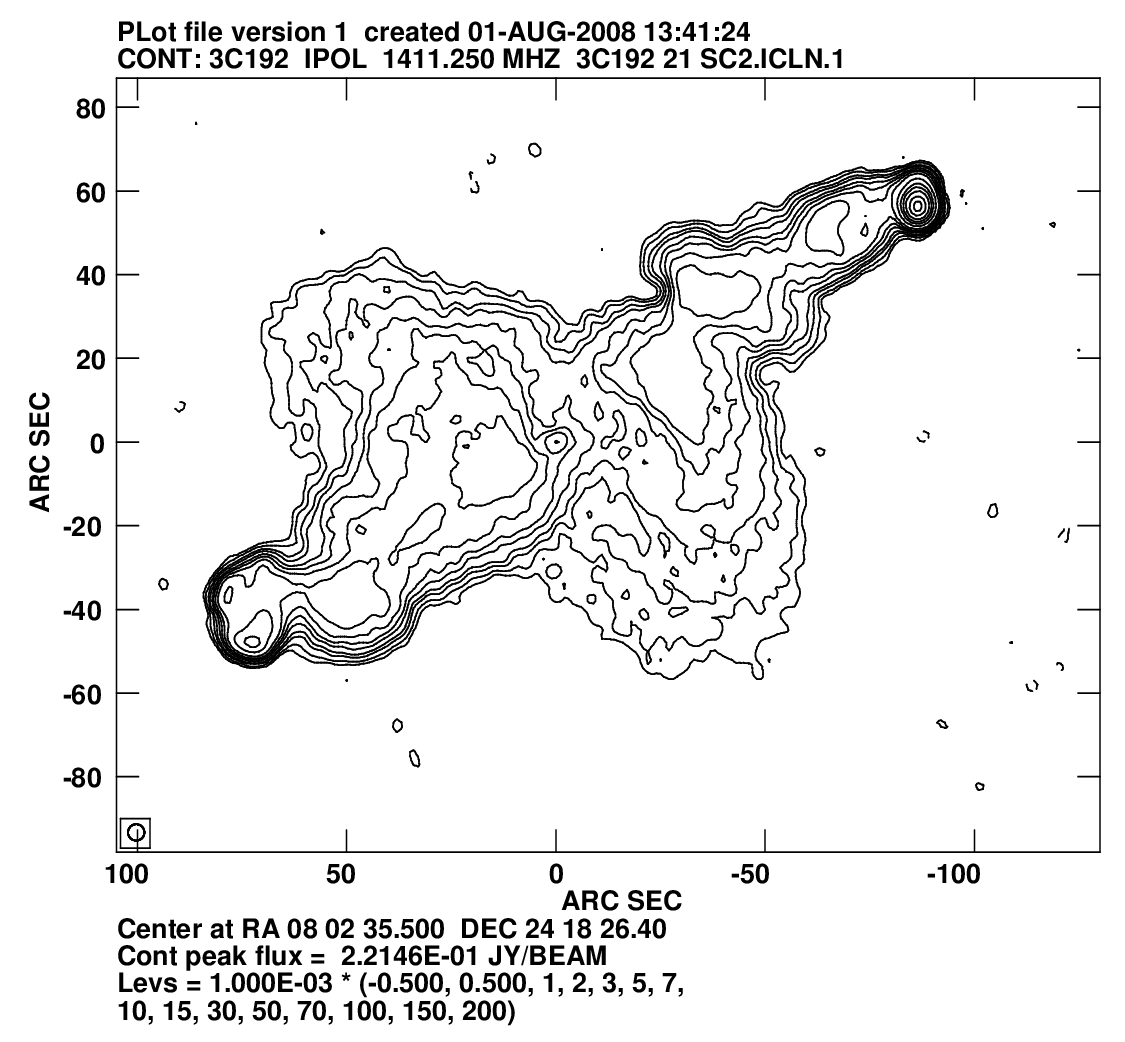}                                   
\caption{\label{fig1a}VLA image of 3C 192 at 1.4 GHz. The HPBW is 
3.9$^\prime\prime$}
\end{minipage}\hspace{2pc}%
\begin{minipage}{14pc}
\includegraphics[width=14pc]{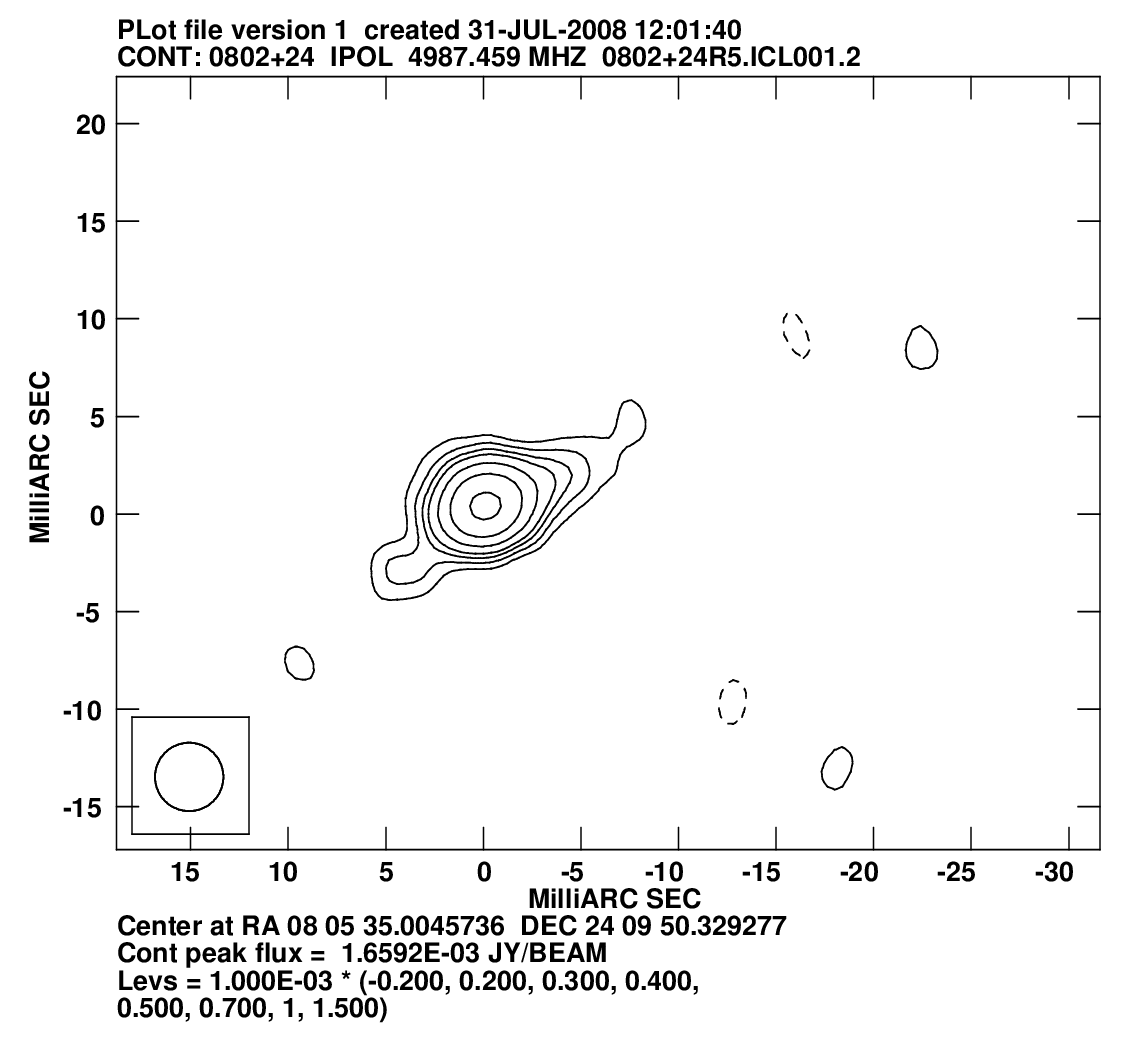}\hspace{2pc}%
\caption{\label{fig1b}VLBA image at 5 GHz of 3C 192. The HPBW is 3.5 mas.}
\end{minipage}       
\end{figure}  

{\bf 0836+29-I -- 4C29.30}.
This source named also B2 0836+29A,
was studied in detail by \cite{bre86}, and recently by
\cite{jam07}. 

\begin{figure}[h]
\begin{minipage}{14pc}
\includegraphics[width=14pc]{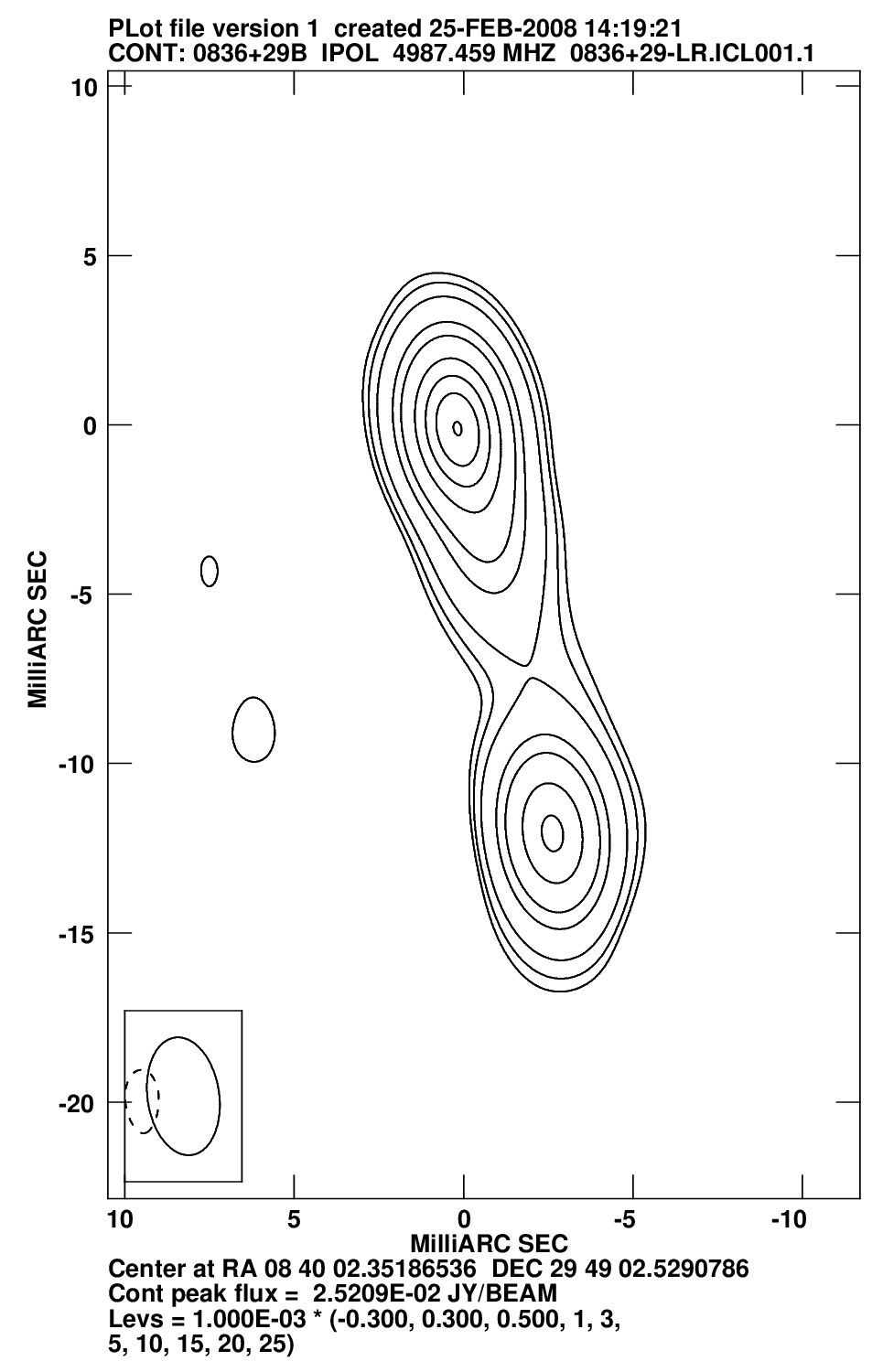}                                   
\caption{\label{fig2a}VLBA image of 0836+29 at 5 GHz. The HPBW is 3.5 $\times$
2.1 mas in PA = 8$^\circ$.}           
\end{minipage}\hspace{2pc}%
\begin{minipage}{14pc}
\includegraphics[width=14pc]{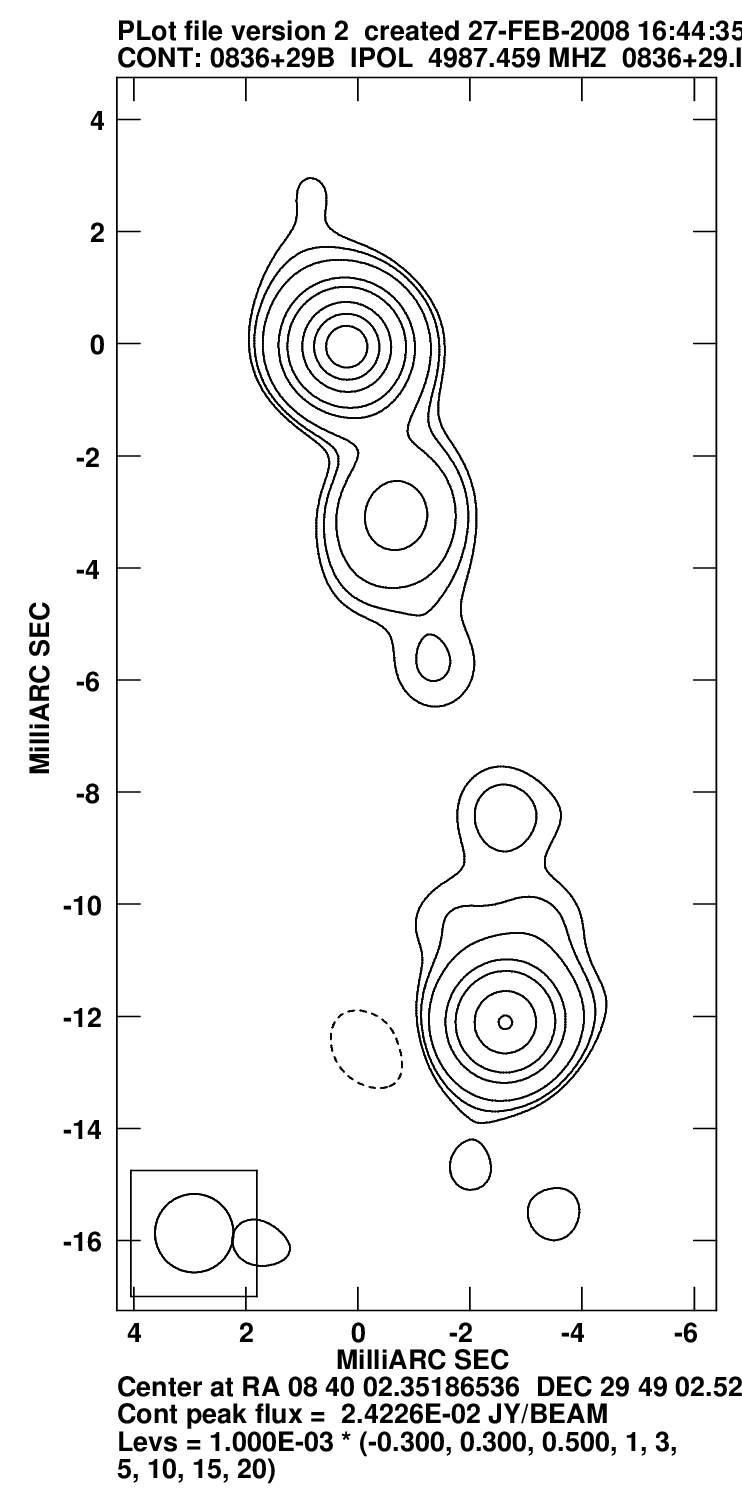}                         
\caption{\label{fig2b}The same as Fig. 3 but with a HPBW = 1.4 mas.}          
\end{minipage}       
\end{figure}  

On the large scale the source shows a clear evidence
of intermittent activity. There is a large scale structure about 9$^\prime$ 
in size,
with an estimate age of about 200 Myr; a more compact structure shows a
central core and two bright spots and an extended emission about 1$^\prime$ 
in size
with an age lower than 100 Myr. The inferred spectral age for the inner
double is 33 Myr \cite{jam07}. 
At higher resolution \cite{bre86} is visible the core source 
(component C2) and a 
one-sided bright jet
about 3'' in size, which terminates with a bright spot (C1).

The high resolution (mas scale) is shown in Fig.3 and Fig.4.  
It looks very similar
to the arcsecond scale by \cite{bre86}. 
At mas resolution we do not
have spectral information, however, we identify the nuclear source with 
the North
component because it is the only unresolved structure, and
because of homogeneity with the arcsecond scale. From the jet to counter-jet
ratio ($>$ 50) we estimate that the orientation angle has to be smaller than
50$^\circ$ and the jet velocity larger than 0.65c. Assuming a high jet
velocity \cite{gio01} with a Lorentz factor 
in the range 3 to 10
and an orientation angle $\sim$ 40$^\circ$, we can derive the intrinsic
jet length and age of single components. In this scenario the first knot after
the core is $\sim$ 15 yrs old and the Southern knot is $\sim$ 70 yr old.
The parsec scale structure is in agreement with a periodic non constant
activity of this source in the small as well as on the large time scale.

We note that \cite{jam07}
discuss on the evidence of a strong outburst in the time range from 1990 to 
2005,
which could be related to the presence of the new component with an estimated
age of 15 yrs.

{\bf B2 1350+31 -- 3C293}.
This peculiar source has been studied in detail in the radio and optical
bands. Recent results are discussed in \cite{bes04}, 
and \cite{flo06},
where the source structure is presented from the sub-arcsec to the
arcmin scale. In \cite{gio05}, we presented a VLBI image at 5 GHz
where a possible symmetric two-sided jet structure is present at mas
resolution. Because of the complex structure of the inner structure we observed
again this source with the VLBA at 1.7 GHz in phase reference mode to properly
map the inner structure. Moreover we reduced VLA data obtained including also
the VLBA-Pt telescope to increase the VLA angular resolution and to study the
connection between the sub-arcsec and the arcsec structure.
In Fig.5 we show the VLBA image at 1.6 GHz where the nuclear source
and a two-sided symmetric emission is seen.

\begin{figure}[h]                                                             
\includegraphics[width=14pc]{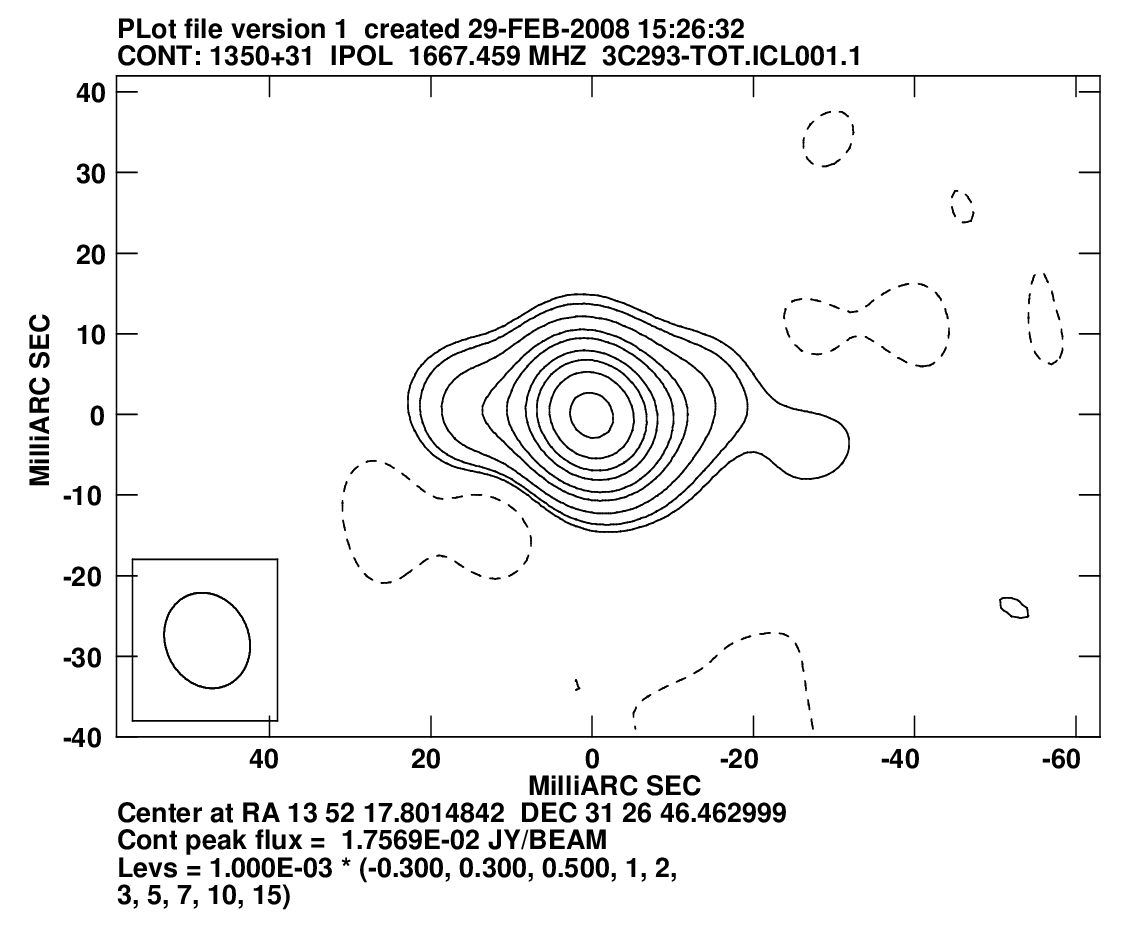}\hspace{2pc}%
\begin{minipage}[b]{14pc}\caption{\label{fig4} VLBA image at 1.6 GHz of 3C 293.
The HPBW is 12 $\times$ 10 mas in PA 24$^\circ$.}
\end{minipage}                                                               
\end{figure} 

In Fig. 6 we show
the VLA and VLBA-Pt image: we have a symmetric structure with 
two symmetric jets. 
The East jet is slightly brighter, in agreement
with the indication discussed in \cite{bes04} that the East jet is
ponting toward us. 
However, the high symmetry
of the VLBA jet suggests that this source is oriented at a large
angle ($\sim$ 75$^\circ$) with respect to the line of sight. 
Therefore, the 
change in the jet PA from the sub-arcsecond to the large scale structure 
(see Figs. 6 and 7) is real and not enhanced by projection effects.

The source brightness between these two structures suggest a two
phase activity: the source emission after some time re-started with
a higher activity with respect to before.
However the jet direction appears to
be constant in between the two activity epochs. In Fig. 6 is already visible
the bending of the the jets in the direction of the kpc scale structure
shown in Fig. 7.
This large change
is not due to a different direction in the restarting nuclear activity but 
it looks constant in time and it is likely produced by the jet interaction
with a rotating disk as discussed in \cite{bre86}.

\begin{figure}[h]
\begin{minipage}{14pc}
\includegraphics[width=14pc]{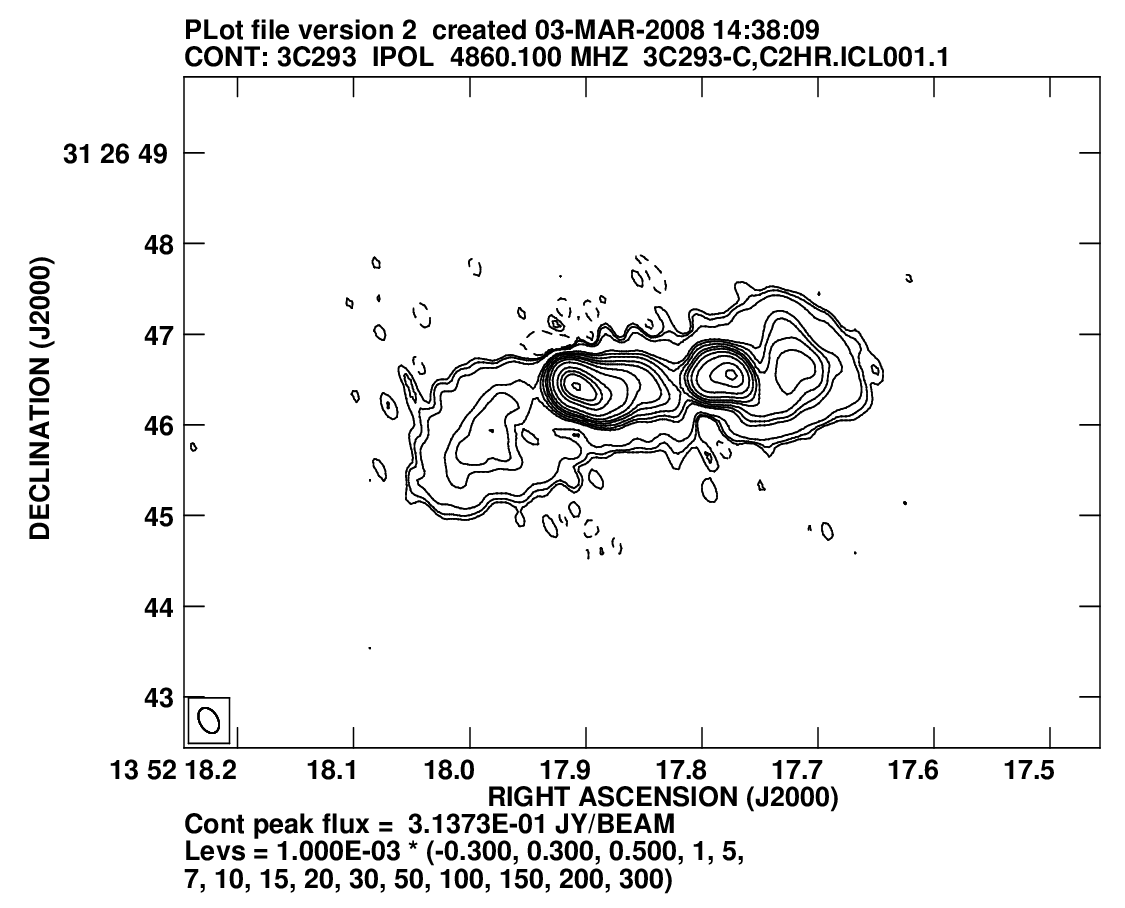}                                   
\caption{\label{fig6}VLA and VLBA-Pt image of 3C293 at 5 GHz. The HPBW is 
0.2 $\times$ 0.3 arcsec in PA 30$^\circ$.}           
\end{minipage}\hspace{2pc}%
\begin{minipage}{14pc}
\includegraphics[width=14pc]{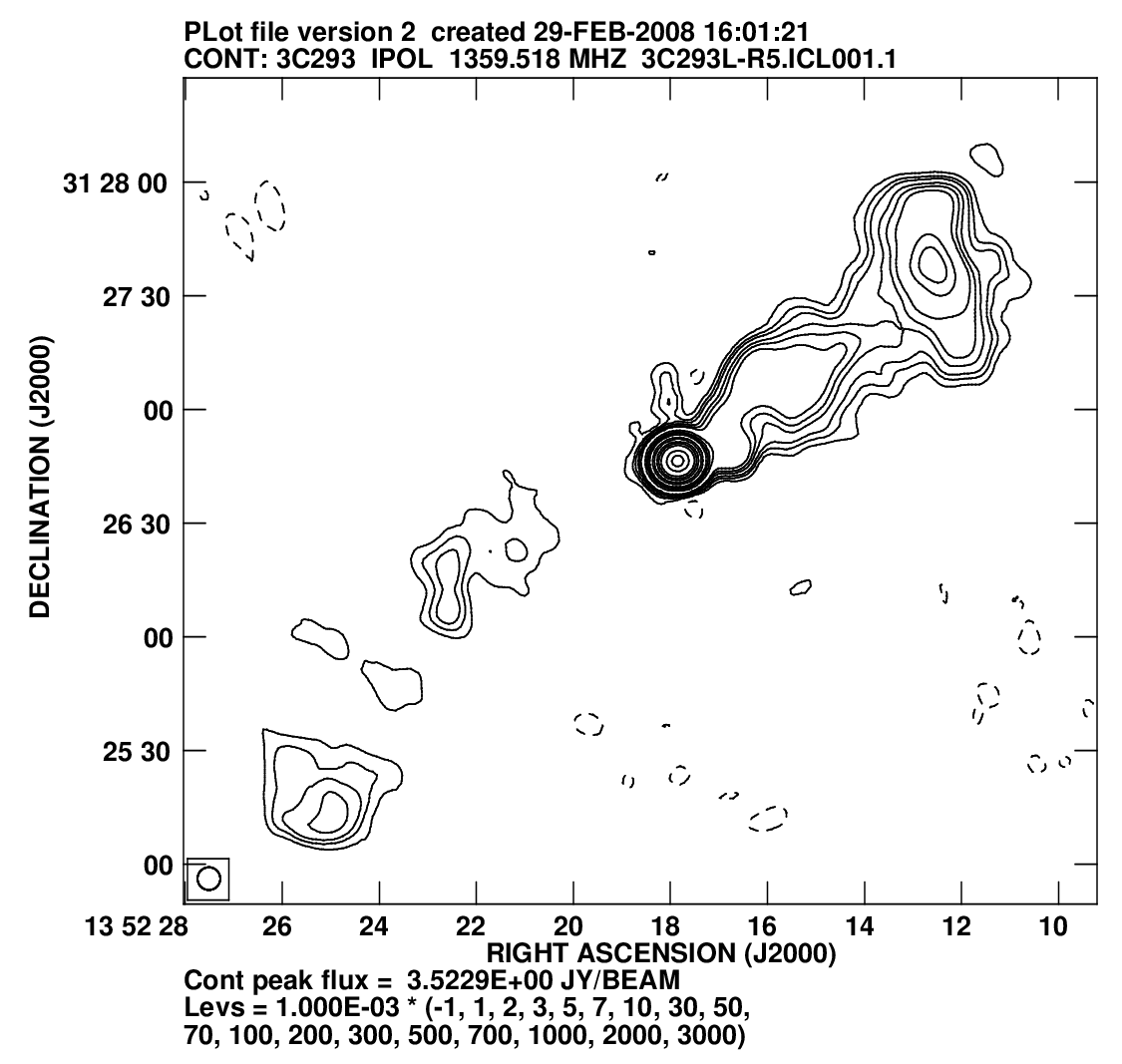}                         
\caption{\label{fig7}VLA image at 1.4 GHz of 3C 293. The HPBW is 6 arcsec.}
\end{minipage}       
\end{figure}  

\section{Conclusions}

Detailed VLBI images of 76 sources of the BCS, a complete sample selected at
low frequency, show the presence of a large number of two-sided jet structures
($\sim$ 22\%) in agreement with a random orientation of radio galaxies and a 
high jet velocity ($\sim$ 0.9c).

Sources with a two-sided jet structure show a faint nuclear emission 
(e.g., 3C 192) and are
identified with galaxies with no broad emission line, in agreement with
prevision of unified models.

In most case the parsec and kiloparsec scale jet structure are aligned
with the main jet on the same side with respect to the nuclear emission, in 
agreement with a continuity of jet properties from the pc to the kpc
scale and no large intrinsic change in the jet position angles.

A few sources show evidence of a restarted activity (e.g. 0836+29-I). The 
recent activity is at the same PA of the older structure. 

In 3C 293 we find
a Z-shaped structure probably due to the
strong interaction of jets with a rotating disk.

\subsection{Acknowledgments}                                       
The National Radio Astronomy
Observatory is operated
by Associated Universities, Inc., under cooperative agreement with the
National Science Foundation.


\section*{References}                                                         

\begin{thebibliography}{13}                                                    

\bibitem{bau88}
Baum S, Heckman T M, Bridle A, van Breugel W J M and Miley G K 
1988 {\it A\&AS} {\bf 68}, 643

\bibitem{bau92}
Baum S A, Heckman T M and van Breugel W 1992, {\it ApJ} {\bf 389}, 208

\bibitem{bes04}
Beswick R J, Peck A B, Taylor G B and Giovannini G 2004, {\it MNRAS} 
{\bf 352}, 49

\bibitem{bre86}
van Breugel W J M, Heckman T M, Miley G K and Filippenko A V 1986,
{\it ApJ} {\bf 311}, 58

\bibitem{fer84}
Feretti L, Giovannini G, Gregorini L, Parma P and Zamorani G 1984, {\it 
A\&A} {\bf 139}, 55

\bibitem{flo06}
Floyd D J E, Perlman E, Leahy J P, Beswick R J, Jackson N J, 
Sparks W B, Axon D J and O'Dea C P 2006 {\it ApJ} {\bf 639}, 23

\bibitem{gio90}
Giovannini G, Feretti L, Gregorini L and Parma P 1988, {\it A\&A} {\bf 199},
73

\bibitem{gio01}
Giovannini G, Cotton W D, Feretti L, Lara L and Venturi T 2001, {\it ApJ}
{\bf 552}, 508

\bibitem{gio05}
Giovannini G, Taylor G B, Feretti L, Cotton W D, Lara L and 
Venturi T 2005, {\it ApJ} {\bf 618}, 635

\bibitem{jam07}
Jamrozy M, Konar C, Saikia D J, Stawarz L, Mack K-H and Siemiginowska A
2007 {\it MNRAS} {\bf 378}, 581

\bibitem{har98}
Hardcastle M J, Alexander P, Pooley G G and Riley J M 1998, 
{\it MNRAS} {\bf 296}, 445

\bibitem{pea88}
Pearson T J and Readhead A C S 1988, {\it ApJ} {\bf 328}, 114                                              

\bibitem{pol95}
Polatidis A G, Wilkinson P N, Xu W, Readhead A C S, Pearson T J, 
Taylor G B and Vermeulen R C 1995, {\it ApJS} {\bf 98}, 1

\bibitem{tay94}
Taylor G B, Vermeulen R C, Pearson T J, Readhead A C S,
 Henstock D R, Browne I W A and Wilkinson P N 1994, {\it ApJS} 
{\bf 95}, 345

\end{thebibliography}
\end{document}